\documentstyle[11pt,psfig,twoside]{article}

\begin{document}

\newcommand{\dd}{\,{\rm d}}
\newcommand{\ie}{{\it i.e.},\,}
\newcommand{\etal}{{\it et al.\ }}
\newcommand{\eg}{{\it e.g.},\,}
\newcommand{\cf}{{\it cf.\ }}
\newcommand{\vs}{{\it vs.\ }}
\newcommand{\zdot}{\makebox[0pt][l]{.}}
\newcommand{\up}[1]{\ifmmode^{\rm #1}\else$^{\rm #1}$\fi}
\newcommand{\dn}[1]{\ifmmode_{\rm #1}\else$_{\rm #1}$\fi}
\newcommand{\upd}{\up{d}}
\newcommand{\uph}{\up{h}}
\newcommand{\upm}{\up{m}}
\newcommand{\ups}{\up{s}}
\newcommand{\arcd}{\ifmmode^{\circ}\else$^{\circ}$\fi}
\newcommand{\arcm}{\ifmmode{'}\else$'$\fi}
\newcommand{\arcs}{\ifmmode{''}\else$''$\fi}
\newcommand{\MS}{{\rm M}\ifmmode_{\odot}\else$_{\odot}$\fi}
\newcommand{\RS}{{\rm R}\ifmmode_{\odot}\else$_{\odot}$\fi}
\newcommand{\LS}{{\rm L}\ifmmode_{\odot}\else$_{\odot}$\fi}

\newcommand{\Abstract}[2]{{\footnotesize\begin{center}ABSTRACT\end{center}
\vspace{1mm}\par#1\par
\noindent
{\bf Key words:~~}{\it #2}}}

\newcommand{\TabCap}[2]{\begin{center}\parbox[t]{#1}{\begin{center}
  \small {\spaceskip 2pt plus 1pt minus 1pt T a b l e}
  \refstepcounter{table}\thetable \\[2mm]
  \footnotesize #2 \end{center}}\end{center}}

\newcommand{\TableSep}[2]{\begin{table}[p]\vspace{#1}
\TabCap{#2}\end{table}}

\newcommand{\TableFont}{\footnotesize}
\newcommand{\TableFontIt}{\ttit}
\newcommand{\SetTableFont}[1]{\renewcommand{\TableFont}{#1}}

\newcommand{\MakeTable}[4]{\begin{table}[htb]\TabCap{#2}{#3}
  \begin{center} \TableFont \begin{tabular}{#1} #4 
  \end{tabular}\end{center}\end{table}}

\newcommand{\MakeTableSep}[4]{\begin{table}[p]\TabCap{#2}{#3}
  \begin{center} \TableFont \begin{tabular}{#1} #4 
  \end{tabular}\end{center}\end{table}}

\newenvironment{references}%
{
\footnotesize \frenchspacing
\renewcommand{\thesection}{}
\renewcommand{\in}{{\rm in }}
\renewcommand{\AA}{Astron.\ Astrophys.}
\newcommand{\AAS}{Astron.~Astrophys.~Suppl.~Ser.}
\newcommand{\ApJ}{Astrophys.\ J.}
\newcommand{\ApJS}{Astrophys.\ J.~Suppl.~Ser.}
\newcommand{\ApJL}{Astrophys.\ J.~Letters}
\newcommand{\AJ}{Astron.\ J.}
\newcommand{\IBVS}{IBVS}
\newcommand{\PASP}{P.A.S.P.}
\newcommand{\Acta}{Acta Astron.}
\newcommand{\MNRAS}{MNRAS}
\renewcommand{\and}{{\rm and }}
\section{{\rm REFERENCES}}
\sloppy \hyphenpenalty10000
\begin{list}{}{\leftmargin1cm\listparindent-1cm
\itemindent\listparindent\parsep0pt\itemsep0pt}}%
{\end{list}\vspace{2mm}}

\def\TYLDA{~}
\newlength{\DW}
\settowidth{\DW}{0}
\newcommand{\dw}{\hspace{\DW}}

\newcommand{\refitem}[5]{\item[]{#1} #2%
\def\REFARG{#3}\ifx\REFARG\TYLDA\else, {\it#3}\fi
\def\REFARG{#4}\ifx\REFARG\TYLDA\else, {\bf#4}\fi
\def\REFARG{#5}\ifx\REFARG\TYLDA\else, {#5}\fi.}

\newcommand{\Section}[1]{\section{#1}}
\newcommand{\Subsection}[1]{\subsection{#1}}
\newcommand{\Acknow}[1]{\par\vspace{5mm}{\bf Acknowledgements.} #1}
\pagestyle{myheadings}

\def\thefootnote{\fnsymbol{footnote}}

\begin{center}
{\Large\bf The Optical Gravitational Lensing Experiment.\\
\vskip3pt
The Catalog of Clusters in the Small Magellanic Cloud
\footnote{Based on observations obtained with the 1.3~m Warsaw telescope at 
the Las Campanas Observatory operated by the Carnegie Institution of 
Washington.}} 
\vskip1cm
{\bf G.~~P~i~e~t~r~z~y~{\'n}~s~k~i$^1$, ~~A.~~U~d~a~l~s~k~i$^1$,~~
M.~~K~u~b~i~a~k$^1$, ~~M.~~S~z~y~m~a~{\'n}~s~k~i$^1$,~~
P.~~W~o~{\'z}~n~i~a~k$^2$,~~ and~~ K.~~{\.Z}~e~b~r~u~{\'n}$^1$}
\vskip5mm
{$^1$Warsaw University Observatory, Al.~Ujazdowskie~4, 00-478~Warszawa,
Poland\\
e-mail: (pietrzyn,udalski,mk,msz,zebrun)@sirius.astrouw.edu.pl\\
\vskip3pt
$^2$ Princeton University Observatory, Princeton, NJ 08544-1001, USA\\ 
e-mail: wozniak@astro.princeton.edu}
\end{center}
\vskip 10mm

\Abstract{We present the catalog of clusters found in the area of ${\approx 
2.4}$ square degree in the central regions of the Small Magellanic Cloud. The 
catalog contains data for 238 clusters, 72 of them are new objects. For each 
cluster equatorial coordinates, radii, approximate number of members, 
cross-identification, finding chart and color magnitude diagrams: ${V-(B-V)}$ 
and ${V-(V-I)}$ are provided. Photometric data for all clusters presented in 
the catalog are available from the OGLE Internet archive.}{Magellanic Clouds 
-- Clusters: general --  Catalogs -- Atlases}

\setcounter{footnote}{0}
\def\thefootnote{\arabic{footnote}}

\Section{Introduction}
The Optical Gravitational Lensing Experiment (OGLE) is a long term observing 
program with the main goal of probing dark matter in the Galaxy with 
microlensing phenomena (Paczy{\'n}ski 1986). The survey started in 1992 and 
its first phase lasted until 1995. About 20 microlensing events (\eg Wo\'zniak 
and Szyma{\'n}ski 1998) including the first ever observed toward the Galactic 
bulge, were detected during this phase. 

From the beginning of 1997 a new phase of the project called OGLE-II has 
started. A major upgrade of the observational capabilities including a new 
dedicated to the project 1.3~m Warsaw telescope has increased the data flow by 
a factor of about 30. New targets were added, namely the Magellanic Clouds and 
Galactic disk fields. Also regularly observed region of the Galactic bulge was 
increased to about 10 square degrees. History of the OGLE project, 
detailed description of the telescope, instrumentation and data pipeline can 
be found in Udalski, Szyma{\'n}ski and Kubiak (1997). 

Huge databases of photometric measurements collected in the course of the 
OGLE-II project may be used to many side projects unrelated to microlensing. 
High quality of collected photometric data, long time baseline of observations 
make the data ideal to address many issues of modern astrophysics. The densest 
stellar regions of the sky observed during the OGLE-II project have been 
rarely observed with modern techniques until now. Thus the OGLE-II 
observations obtained with the standard {\it BVI}-band filters provide a 
unique photometric material to study these objects. 

The main targets of the OGLE-II project include the Large and Small Magellanic 
Clouds. About 2.4 and 4.5 square degree regions covering the central bars of 
the SMC and LMC, respectively, are monitored each night during the observing 
season of each galaxy. Collected data should help to answer many puzzling 
questions concerning these galaxies. For instance, photometry of the red clump 
stars and RR~Lyr variable stars allowed precise determination of distances 
to both galaxies (Udalski \etal 1998a, Udalski 1998). Three-color {\it BVI} 
maps of the SMC have already been released providing precise photometry and 
astrometry for about 2.2 million stars in this galaxy (Udalski \etal 1998b). 
Similar maps for the LMC will follow. Catalogs of thousands variable stars 
will be released in the near future. 

In this paper we present another side project of the OGLE-II survey -- the 
catalog of clusters in the SMC which is the first part of the series of 
catalogs of stellar clusters in the Magellanic Clouds. 

Observations of cluster systems in other galaxies are very important. They can 
answer many questions concerning the nature of stellar clusters and their 
formation and disruption processes. They also provide information about 
stellar populations and their evolution in parent galaxies. Due to proximity, 
relatively small extinction and difference in metallicity, the rich system of 
Magellanic Cloud clusters is especially good for such studies. 

Previous searches for clusters in the SMC revealed 734 candidate objects. 220 
clusters were cataloged by 1975 (Hodge and Wright 1977). This atlas comprise 
116, 69, 18, and 86 clusters discovered by Linsday (1958), Kron (1956), 
Westerlund and Glaspey (1971) and Hodge and Wright (1974), respectively. 
Br\"uck (1976) presented additional 168 cluster candidates. Later, Hogde 
(1986) published positions of 213 new SMC clusters. However, many objects from 
this sample were flagged as doubtful or questionable. A revised and extended 
catalog comprising known and additional 133 clusters in the SMC was published 
by Bica and Schmitt (1995). All these catalogs are based on searches performed 
manually. Visual examination of images is very often subjective and may lead 
to spurious detections. In fact many clusters discovered in the past were not 
confirmed by Bica and Schmitt (1995). 

For a long time accurate positions of SMC clusters were not available. In 1991 
Welch (1991) published the first list of accurate coordinates, but only for 
203 clusters from the SMC. The catalog of Bica and Schmitt (1995) contains 
more information including good quality coordinates of SMC clusters. 
Unfortunately, precise photometric data exist only for handful of clusters in 
the SMC. This galaxy has been neglected photometrically for years. 

The main purpose of this paper is to remove these deficiencies. We present a 
catalog of 238 clusters detected in the OGLE-II SMC fields. The majority of 
clusters were detected with the algorithmic search. Additionally we include a 
list of probable clusters discovered during visual inspection of the fields. 
For each object we provided its equatorial coordinates, size and photometry, 
namely ${V-(B-V)}$ and ${V-(V-I)}$ color-magnitude diagrams (CMDs). All data 
are available to the astronomical community from the OGLE Internet archive. 

\Section{The Photometric Data}
All observations presented in this paper were collected during the second 
phase of the OGLE microlensing survey with the 1.3~m Warsaw telescope at the 
Las Campanas Observatory, Chile, which is operated by the Carnegie Institution 
of Washington. The details of the system can be found in Udalski, 
Szyma{\'n}ski and Kubiak (1997). 

Photometric data of the SMC come from the three-color, {\it BVI} maps of the 
SMC (Udalski \etal 1998b). Full description of the reduction techniques, tests 
of data quality of these maps can be found in Udalski \etal (1998b). The maps 
provide precise {\it BVI} photometry of about 2.2 million stars in the SMC. The 
magnitudes were obtained from tens/hundreds of individual measurements. 
Accuracy of the zero points of the OGLE photometry in the SMC is about 
0.01~mag. Complete photometry is obtained down to ${V\approx21.5}$~mag. 

Eleven driftscan fields, covering large part of the central bar of the SMC 
were observed: SMC$\_$SC1--SMC$\_$SC11. Coordinates of the center of each 
field as well as the SMC map with contours of observed fields can be found in 
Udalski \etal (1998b) (see their Fig.~1 and Table~1). 

\Section{Search for Clusters in the SMC}
Because of the subjective nature of visual search, the large area covered by 
observations and crowded stellar background in the observed parts of the SMC 
we decided to perform an automatic, algorithmic search for stellar clusters. 
We applied similar technique as Zaritsky, Harris and Thompson (1997) in their 
search for clusters in selected field of the LMC. 

First, each of our eleven driftscans was divided into square boxes and stars 
having both {\it V} and {\it I}-band photometry were counted inside each of 
them. In order to account for different sizes of potential clusters we used 
boxes of ${20\times20}$, ${30\times30}$ and ${50\times50}$ pixels (${8.2\times 
8.2}$, ${12.3\times12.3}$ and ${20.5\times20.5}$~arcsec, respectively). 
Density maps obtained with boxes ${20\times20}$ pixels are well suited for 
search for small clusters, while those made with grid ${50\times50}$ pixels 
are much better for larger ones. The set of density images obtained for scan 
SMC$\_$SC6 is shown in Fig.~1. The intensity of pixels reflects the number of 
stars counted in a given box.

Next, the procedure called "unsharp masking" was employed to remove background 
variations from our density maps. For each density map two images were 
produced. They were obtained by convolving density maps with two dimensional 
Gaussian with ${\sigma_{x}=\sigma_{y}=2.0}$ and ${\sigma_{x}=\sigma_{y}=6.0}$ 
pixels, respectively. The images obtained as a difference between the latter 
and former (\ie unsharp masked images) were then searched for significant 
pixels. The unsharp masked images of stellar density of scan SMC$\_$SC6 are 
presented in Fig.~2. The suspected clusters were selected as concentrations of 
at least four pixels exceeding a given threshold above the mean. To obtain 
information on reality of a potential cluster the procedure was repeated with 
three different thresholds of 2, 3, and 4 standard deviations of all pixels of 
the map. The preliminary positions of candidates for cluster were obtained as 
the mean of the significant pixels in the group. 

Finally, to avoid false detections each candidate was carefully examined. 
Numerous detections were due to bright overexposed stars. Groups of pixels 
extending over the threshold of 4 and $3\sigma$ were classified as "certain" 
and "very probable" clusters, respectively. Because the threshold of  
$2\sigma$ above the mean triggered many spurious detection caused by density 
fluctuations, all detection with this threshold were further carefully 
examined visually. Those resembling clusters were classified as "probable". 

In total, 76, 33 and 52 of certain, very probable and probable clusters, 
respectively, were detected in eleven driftscan fields of the SMC.

During further visual examination of our images 77 additional candidates for 
clusters were found. Because they are relatively faint and small our algorithm 
did not trigger them, and therefore we list them separately. 

We compared coordinates of our objects with that from the list of clusters 
with accurate positions published by Bica and Schmitt (1995). There are 166 
clusters common to both catalogs. 

\Section{Atlas of Clusters in the SMC}
\Subsection{Equatorial Coordinates of Clusters}
The equatorial coordinates (J2000) of each cluster were estimated in the 
following way. We started with crude coordinates obtained as the mean of 
positions of significant pixels from unsharp masked density maps (see 
Section~3). Then, coordinates of the cluster center were iteratively obtained 
by deriving average position of stars inside the circle of 100 pixels (41 
arcsec). After a few iterations its error was smaller than 0.1 pixel. To test 
consistency of the procedure we shifted the obtained position and repeated the 
procedure. Ten such tests were performed and ten different centers were 
obtained for each cluster. The mean value was adopted as the final position of 
the center of the cluster. The standard deviation of position depended on 
richness of the cluster. For large, populous clusters dispersion was about 5 
arcsec while for small poorly populated almost two times larger. The 
transformation from the pixel system to equatorial coordinates is described in 
detail in Udalski \etal (1998b) 

\Subsection{Angular Size of Clusters}
In order to estimate the angular extend of clusters we performed counts of 
stars in annuli around center of each cluster. Our data allow to obtain counts 
far away from the cluster center and to define clearly the stellar background. 
Density profiles obtained in this way were then visually examined. 

Typical density profile of a cluster shows some fluctuations after the main 
drop of the stellar density. Such a behavior is most likely caused by extended 
halos around clusters. 

Unfortunately the stellar background in the observed SMC fields is relatively 
high and halo regions of clusters contain usually many field stars. In several 
cases the cluster CMDs obtained for stars within the radius up to the end of 
density fluctuations were contaminated and general features of cluster CMD 
were severely affected. Therefore we decided to define also the cluster core 
radius as the distance from the cluster center to the major drop of the 
stellar density. These radii represent sizes of cluster cores rather than 
entire cluster and define regions where cluster stars are more frequent than 
field ones. 

\Subsection{The Catalog}
Table~1 shows acronyms used in our catalog. Table~2 contains information about 
each cluster found in the automatic way. In columns 1, 2 and 3 we list 
proposed name and equatorial coordinates RA and DEC (J2000), respectively. The 
name consists of the suffix OGLE-CL- with object name and four digit number, 
\eg OGLE-CL-SMC0001. In column 4 we provide information about the stellar 
density of cluster relative to the background. Abbreviations "c", "vp" and "p" 
have the following meaning: certain, very probable and probable, respectively 
(Section~3). In columns 5, 6, 7 and 8 we provide the radius, core radius, 
approximate number of cluster members and cross-identification. 
Remark~(1) in column~1 indicates that the cluster contains bright star, (2) -- 
the cluster is located near the edge of the frame, and (3)~--~it is a member 
of cluster pair. Remark (4) indicates that due to severe crowding the 
photometric data were filtered for CMD construction. Only stars with standard 
error of the {\it I}-band magnitude smaller than two times of the maximum 
standard error of non-variable stars of a given magnitude were used. 

\renewcommand{\arraystretch}{1.1}
\MakeTable{|llr|}{8cm}{SMC catalogs}
{\hline
Acronym & Reference & Entries \\
\hline
K    & Kron 1956             & 27 \\
L    & Lindsay 1958          & 35 \\
L61  & Lindsay 1961          & 8 \\
HW   & Hodge and Wright 1974 & 3 \\
B    & Br{\"u}ck 1976        & 37 \\
H86  & Hodge 1986            & 61 \\
BS   & Bica and Smith 1995   & 23 \\
OGLE & this paper            & 72 \\
\hline
}

\renewcommand{\arraystretch}{1}
\MakeTableSep{|l|c|c|c|c|c|c|c|}{12.5cm}{SMC clusters}{
\hline
\multicolumn{1}{|c|}{Name}&
\multicolumn{1}{|c|}{$\alpha_{2000}$}&
\multicolumn{1}{|c|}{$\delta_{2000}$}&
\multicolumn{1}{|c|}{class}&
\multicolumn{1}{|c|}{$R$}&
\multicolumn{1}{|c|}{$R_{\rm core}$ }&
\multicolumn{1}{|c|}{N} &
\multicolumn{1}{|c|}{Cross-}\\
\multicolumn{1}{|c|} {OGLE-CL-} & & & & 
\multicolumn{1}{|c|}{[\arcs]} & {[\arcs]} & &-identifications \\
\hline
SMC0001 & 0\uph36\upm38\zdot\ups34 & $-73\arcd05\arcm09\zdot\arcs2$ & vp &  16 &  12 &  11 & H86-35\\
SMC0002 & 0\uph37\upm33\zdot\ups06 & $-73\arcd36\arcm42\zdot\arcs6$ & c &  47 &  31 &  127 & HW11  \\
SMC0003 & 0\uph37\upm42\zdot\ups24 & $-73\arcd54\arcm29\zdot\arcs5$ & c &  42 &  42 &  223 & L19    \\
SMC0004$^{(1)}$ & 0\uph37\upm44\zdot\ups03 & $-73\arcd12\arcm39\zdot\arcs8$ & vp &  36 &  36 &  30 & B10\\
SMC0005$^{(2)}$ & 0\uph39\upm21\zdot\ups63 & $-73\arcd15\arcm28\zdot\arcs4$ & vp &  22 &  11 &  16 & OGLE\\
SMC0006 & 0\uph39\upm33\zdot\ups12 & $-73\arcd10\arcm37\zdot\arcs2$ & p &  24 &  16 &   15 & OGLE        \\
SMC0007 & 0\uph40\upm30\zdot\ups49 & $-73\arcd13\arcm50\zdot\arcs2$ & p &  28 &  12 &   17 & OGLE        \\
SMC0008 & 0\uph40\upm30\zdot\ups54 & $-73\arcd24\arcm10\zdot\arcs4$ & c &  43 &  24 &  256 & NGC220,K18,L22 \\
SMC0009 & 0\uph40\upm44\zdot\ups11 & $-73\arcd23\arcm00\zdot\arcs2$ & c &  36 &  18 &  128 & NGC222,K19,L24 \\
SMC0010 & 0\uph40\upm47\zdot\ups58 & $-73\arcd05\arcm16\zdot\arcs7$ & vp &  12 &  12 &  15 & H86-62         \\
SMC0011 & 0\uph41\upm06\zdot\ups16 & $-73\arcd21\arcm07\zdot\arcs1$ & c &  36 &  24 &  131 & NGC231,K20,L25 \\
SMC0012 & 0\uph41\upm23\zdot\ups78 & $-72\arcd53\arcm27\zdot\arcs1$ & c &  61 &  36 &  240 & K21,L27        \\
SMC0013 & 0\uph42\upm22\zdot\ups37 & $-73\arcd44\arcm03\zdot\arcs1$ & c &  23 &  16 &   30 & HW16           \\
SMC0014$^{(2)}$ & 0\uph42\upm27\zdot\ups89 & $-73\arcd32\arcm49\zdot\arcs9$ & p &  24 &  12 &   17 & OGLE   \\
SMC0015 & 0\uph42\upm54\zdot\ups13 & $-73\arcd17\arcm37\zdot\arcs0$ & c &  30 &  11 &   23 & OGLE           \\
SMC0016 & 0\uph42\upm58\zdot\ups46 & $-73\arcd10\arcm07\zdot\arcs2$ & c &  42 &  29 &  130 & BS16           \\
SMC0017$^{(3)}$ & 0\uph43\upm32\zdot\ups74 & $-73\arcd26\arcm25\zdot\arcs4$ & c &  26 &  20 &  261 & NGC241,K22,L29 \\
SMC0018$^{(3)}$ & 0\uph43\upm37\zdot\ups57 & $-73\arcd26\arcm37\zdot\arcs9$ & c &  20 &  14 &  110 & NGC242,K22,L29 \\
SMC0019 & 0\uph43\upm37\zdot\ups59 & $-72\arcd57\arcm30\zdot\arcs9$ & p &  12 &  10 &   30 & B31                    \\
SMC0020$^{(3)}$ & 0\uph43\upm37\zdot\ups89 & $-72\arcd58\arcm48\zdot\arcs3$ & p &   9 &   9 &    6 & BS20           \\
SMC0021$^{(3)}$ & 0\uph43\upm44\zdot\ups40 & $-72\arcd58\arcm35\zdot\arcs6$ & c &  36 &  18 &   41 & H86-70         \\
SMC0022 & 0\uph43\upm57\zdot\ups94 & $-73\arcd09\arcm07\zdot\arcs5$ & p &  22 &  12 &   17 & OGLE                   \\
SMC0023 & 0\uph44\upm12\zdot\ups57 & $-73\arcd37\arcm08\zdot\arcs1$ & p &  18 &  12 &   27 & B33                    \\
SMC0024 & 0\uph45\upm01\zdot\ups22 & $-72\arcd55\arcm17\zdot\arcs3$ & vp &  57 &  30 &  92 & OGLE                   \\
SMC0025 & 0\uph45\upm13\zdot\ups88 & $-73\arcd13\arcm09\zdot\arcs2$ & vp &  15 &  14 &  65 & H86-74                 \\
SMC0026$^{(1)(3)(4)}$& 0\uph45\upm24\zdot\ups07 & $-73\arcd22\arcm44\zdot\arcs4$ & c &  49 &  24 &   20 & NGC248,L61-67\\
SMC0027$^{(3)}$ & 0\uph45\upm25\zdot\ups66 & $-73\arcd28\arcm52\zdot\arcs7$ & c &  36 &  24 &   79 & B39                    \\
SMC0028 & 0\uph45\upm27\zdot\ups92 & $-72\arcd49\arcm09\zdot\arcs7$ & vp &  12 &   8 &  16 &OGLE                         \\
SMC0029$^{(2)}$ & 0\uph45\upm29\zdot\ups77 & $-73\arcd04\arcm40\zdot\arcs8$ & c &  22 &  18 &   18 & L61-68                \\
SMC0030 & 0\uph45\upm32\zdot\ups86 & $-73\arcd06\arcm26\zdot\arcs7$ & p &  22 &  12 &   14 & OGLE                           \\
SMC0031 & 0\uph45\upm51\zdot\ups37 & $-72\arcd50\arcm25\zdot\arcs4$ & p &  20 &  12 &   42 & B36                            \\
SMC0032$^{(1)(4)}$ & 0\uph45\upm54\zdot\ups33 & $-73\arcd30\arcm24\zdot\arcs2$ & c &  30 &  19 &  142 & NGC256,K23,L30       \\
SMC0033$^{(3)}$ & 0\uph46\upm12\zdot\ups26 & $-73\arcd23\arcm34\zdot\arcs0$ & p &  18 &  12 &   10 & H86-78               \\
SMC0034 & 0\uph46\upm32\zdot\ups96 & $-73\arcd05\arcm55\zdot\arcs3$ & p &  30 &  12 &   16 & NGC261,B42           \\
SMC0035$^{(3)}$ & 0\uph46\upm33\zdot\ups72 & $-72\arcd46\arcm25\zdot\arcs9$ & p &  14 &  13 &    8 & H86-83         \\
SMC0036 & 0\uph46\upm34\zdot\ups98 & $-72\arcd44\arcm32\zdot\arcs1$ & c &  24 &  20 &   86 & L31                  \\
SMC0037 & 0\uph46\upm41\zdot\ups03 & $-72\arcd59\arcm59\zdot\arcs6$ & p &  24 &  11 &   31 & OGLE                   \\
SMC0038 & 0\uph47\upm06\zdot\ups15 & $-73\arcd15\arcm24\zdot\arcs9$ & p &  21 &  18 &   48 & H86-89               \\
SMC0039$^{(1)(4)}$ & 0\uph47\upm11\zdot\ups61 & $-73\arcd28\arcm38\zdot\arcs1$ & c &  49 &  30 &  301 & NGC265,K24,L34 \\
SMC0040 & 0\uph47\upm01\zdot\ups18 & $-73\arcd23\arcm34\zdot\arcs7$ & vp &  16 &  14 &  13 & H86-86             \\
SMC0041 & 0\uph47\upm24\zdot\ups84 & $-72\arcd50\arcm26\zdot\arcs7$ & vp &  13 &  13 &  41 & L33                \\
SMC0042$^{(3)}$ & 0\uph47\upm49\zdot\ups72 & $-73\arcd28\arcm42\zdot\arcs2$ & d &  16 &  14 &    8 & BS35       \\
SMC0043 & 0\uph47\upm52\zdot\ups38 & $-73\arcd13\arcm20\zdot\arcs3$ & c &  22 &  16 &   74 & H86-97             \\
\hline
}

\setcounter{table}{1}
\MakeTableSep{|l|c|c|c|c|c|c|c|}{12.5cm}{Continued}{
\hline
\multicolumn{1}{|c|}{Name}&
\multicolumn{1}{|c|}{$\alpha_{2000}$}&
\multicolumn{1}{|c|}{$\delta_{2000}$}&
\multicolumn{1}{|c|}{class}&
\multicolumn{1}{|c|}{$R$}&
\multicolumn{1}{|c|}{$R_{\rm core}$ }&
\multicolumn{1}{|c|}{N} &
\multicolumn{1}{|c|}{Cross-}\\
\multicolumn{1}{|c|} {OGLE-CL-} & & & &
\multicolumn{1}{|c|}{[\arcs]} & {[\arcs]} & &-identifications\\
\hline
SMC0044 & 0\uph47\upm54\zdot\ups50 & $-72\arcd57\arcm20\zdot\arcs3$ & vp &  30 &  24 &  50 & H86-98            \\
SMC0045$^{(3)}$ & 0\uph48\upm00\zdot\ups68 & $-73\arcd29\arcm10\zdot\arcs3$ & c &  35 &  32 &  173 & K25,L35   \\
SMC0046$^{(4)}$ & 0\uph48\upm21\zdot\ups23 & $-73\arcd31\arcm49\zdot\arcs1$ & c &  29 &  18 &  218 & NGC269,K26,L37    \\
SMC0047$^{(2)}$ & 0\uph48\upm28\zdot\ups14 & $-72\arcd59\arcm00\zdot\arcs3$ & vp &  36 &  24 &   23 &OGLE      \\
SMC0048 & 0\uph48\upm33\zdot\ups23 & $-73\arcd18\arcm25\zdot\arcs1$ & c &  49 &  24 &  119 & B47               \\
SMC0049 & 0\uph48\upm37\zdot\ups47 & $-73\arcd24\arcm53\zdot\arcs2$ & vp &  38 &  24 &   64 & B48              \\
SMC0050 & 0\uph48\upm59\zdot\ups02 & $-73\arcd09\arcm03\zdot\arcs8$ & p &  14 &  12 &   22 & OGLE              \\
SMC0051 & 0\uph49\upm04\zdot\ups79 & $-73\arcd03\arcm03\zdot\arcs8$ & p &  18 &  11 &   34 & H86-103           \\
SMC0052$^{(2)}$ & 0\uph49\upm12\zdot\ups21 & $-73\arcd06\arcm30\zdot\arcs7$ & vp &  18 &   8 &   31 & H86-104  \\
SMC0053 & 0\uph49\upm17\zdot\ups52 & $-73\arcd12\arcm41\zdot\arcs8$ & p &  30 &  12 &   20 & OGLE              \\
SMC0054 & 0\uph49\upm17\zdot\ups60 & $-73\arcd22\arcm19\zdot\arcs8$ & c &  27 &  18 &   87 & L39               \\
SMC0055 & 0\uph49\upm21\zdot\ups30 & $-73\arcd11\arcm01\zdot\arcs7$ & p &  18 &  12 &   17 & OGLE              \\
SMC0056 & 0\uph49\upm36\zdot\ups29 & $-72\arcd50\arcm12\zdot\arcs7$ & p &  24 &  12 &    7 & OGLE              \\
SMC0057 & 0\uph49\upm39\zdot\ups75 & $-73\arcd03\arcm31\zdot\arcs7$ & c &  49 &  30 &  150 & B52               \\
SMC0058$^{(3)}$ & 0\uph49\upm45\zdot\ups43 & $-72\arcd51\arcm58\zdot\arcs0$ & c &  36 &  20 &   32 & H86-109   \\
SMC0059 & 0\uph50\upm16\zdot\ups06 & $-73\arcd01\arcm59\zdot\arcs6$ & p &  25 &  25 &   62 & BS45              \\
SMC0060 & 0\uph50\upm21\zdot\ups95 & $-73\arcd23\arcm16\zdot\arcs5$ & c &  36 &  18 &   91 & B55               \\
SMC0061 & 0\uph50\upm00\zdot\ups26 & $-73\arcd15\arcm17\zdot\arcs7$ & vp &  21 &  18 &   37 & H86-107          \\
SMC0062 & 0\uph50\upm28\zdot\ups09 & $-73\arcd12\arcm11\zdot\arcs7$ & c &  18 &  11 &   28 & B54               \\
SMC0063 & 0\uph50\upm36\zdot\ups83 & $-73\arcd03\arcm28\zdot\arcs0$ & p &  30 &  24 &   34 & H86-115           \\
SMC0064 & 0\uph50\upm39\zdot\ups55 & $-72\arcd57\arcm54\zdot\arcs8$ & c &  36 &  19 &  121 & H86-116           \\
SMC0065 & 0\uph50\upm54\zdot\ups62 & $-73\arcd03\arcm26\zdot\arcs9$ & p &  20 &  12 &   30 & OGLE              \\
SMC0066 & 0\uph50\upm55\zdot\ups39 & $-73\arcd12\arcm11\zdot\arcs0$ & p &  17 &  14 &   28 & B56               \\
SMC0067 & 0\uph50\upm55\zdot\ups54 & $-72\arcd43\arcm39\zdot\arcs7$ & c &  42 &  18 &  154 & L41               \\
SMC0068 & 0\uph50\upm56\zdot\ups26 & $-73\arcd17\arcm21\zdot\arcs1$ & p &  55 &  24 &  132 & BS40              \\
SMC0069 & 0\uph51\upm14\zdot\ups13 & $-73\arcd09\arcm41\zdot\arcs5$ & c &  36 &  24 &  228 & NGC290,L42        \\
SMC0070$^{(2)}$ & 0\uph51\upm26\zdot\ups15 & $-73\arcd16\arcm59\zdot\arcs8$ & vp &  14 &  14 &   27 & BS251    \\
SMC0071$^{(2)}$ & 0\uph51\upm31\zdot\ups78 & $-73\arcd00\arcm38\zdot\arcs3$ & c &  32 &  24 &   61 & OGLE      \\
SMC0072 & 0\uph51\upm41\zdot\ups69 & $-73\arcd13\arcm46\zdot\arcs8$ & p &  28 &  18 &   36 & OGLE              \\
SMC0073$^{(3)}$ & 0\uph51\upm44\zdot\ups03 & $-72\arcd50\arcm25\zdot\arcs1$ & c &  42 &  24 &  131 & L61-184   \\
SMC0074 & 0\uph51\upm52\zdot\ups91 & $-72\arcd57\arcm13\zdot\arcs9$ & c &  42 &  27 &  187 & K29,L44           \\
SMC0075 & 0\uph51\upm54\zdot\ups32 & $-73\arcd05\arcm52\zdot\arcs9$ & p &  15 &  14 &   32 & H86-126           \\
SMC0076 & 0\uph52\upm12\zdot\ups47 & $-72\arcd31\arcm51\zdot\arcs2$ & c &  29 &  17 &   60 & H86-129           \\
SMC0077 & 0\uph52\upm13\zdot\ups34 & $-73\arcd00\arcm12\zdot\arcs2$ & p &  18 &  16 &    8 & BS56              \\
SMC0078 & 0\uph52\upm16\zdot\ups56 & $-73\arcd01\arcm04\zdot\arcs0$ & p &  36 &  20 &   41 & H86-130           \\
SMC0079 & 0\uph52\upm16\zdot\ups73 & $-73\arcd22\arcm32\zdot\arcs5$ & vp &  32 &  21 &   53 & H86-125          \\
SMC0080 & 0\uph52\upm31\zdot\ups46 & $-72\arcd37\arcm46\zdot\arcs5$ & p &  12 &   8 &   23 & H86-132           \\
SMC0081 & 0\uph52\upm33\zdot\ups65 & $-72\arcd40\arcm53\zdot\arcs6$ & c &  31 &  18 &   97 & H86-133           \\
SMC0082 & 0\uph52\upm42\zdot\ups12 & $-72\arcd55\arcm31\zdot\arcs6$ & vp &  36 &  17 &   53 & BS60             \\
SMC0083$^{(1)}$ & 0\uph52\upm44\zdot\ups27 & $-72\arcd58\arcm47\zdot\arcs8$ & c &  29 &  18 &   80 & B65       \\
SMC0084$^{(3)}$ & 0\uph52\upm46\zdot\ups69 & $-73\arcd24\arcm25\zdot\arcs4$ & p &  12 &  12 &   28 & BS63      \\
SMC0085 & 0\uph52\upm47\zdot\ups64 & $-72\arcd47\arcm45\zdot\arcs7$ & vp &  25 &  12 &   38 & B66              \\
SMC0086$^{(2)}$ & 0\uph52\upm48\zdot\ups18 & $-72\arcd30\arcm38\zdot\arcs5$ & vp &  20 &  12 &   40 & H86-135  \\
\hline
}

\setcounter{table}{1}
\MakeTableSep{|l|c|c|c|c|c|c|c|}{12.5cm}{Continued}{
\hline
\multicolumn{1}{|c|}{Name}&
\multicolumn{1}{|c|}{$\alpha_{2000}$}&
\multicolumn{1}{|c|}{$\delta_{2000}$}&
\multicolumn{1}{|c|}{class}&
\multicolumn{1}{|c|}{$R$}&
\multicolumn{1}{|c|}{$R_{\rm core}$ }&
\multicolumn{1}{|c|}{N} &
\multicolumn{1}{|c|}{Cross-}\\
\multicolumn{1}{|c|} {OGLE-CL-} & & & &
\multicolumn{1}{|c|}{[\arcs]} & {[\arcs]} & &-identifications\\
\hline
SMC0087$^{(3)}$ & 0\uph52\upm48\zdot\ups99 & $-73\arcd24\arcm43\zdot\arcs3$ & c &  22 &  17 &  123 & B67                    \\
SMC0088 & 0\uph53\upm00\zdot\ups61 & $-72\arcd53\arcm48\zdot\arcs5$ & c &  48 &  30 &  440 & K31,L46                        \\
SMC0089 & 0\uph53\upm05\zdot\ups28 & $-72\arcd37\arcm27\zdot\arcs8$ & c & 118 &  18 &   87 & B69                            \\
SMC0090$^{(4)}$ & 0\uph53\upm05\zdot\ups59 & $-73\arcd22\arcm49\zdot\arcs4$ & c &  38 &  24 &  410 & NGC294,L47           \\
SMC0091 & 0\uph53\upm09\zdot\ups53 & $-72\arcd34\arcm24\zdot\arcs9$ & vp &  36 &  18 &   59 & H86-138               \\
SMC0092$^{(1)}$ & 0\uph53\upm17\zdot\ups90 & $-72\arcd45\arcm59\zdot\arcs5$ & c &  34 &  20 &   93 & B71             \\
SMC0093 & 0\uph53\upm31\zdot\ups29 & $-72\arcd40\arcm04\zdot\arcs2$ & vp &  18 &  18 &   25 & H86-143                 \\
SMC0094$^{(3)}$ & 0\uph53\upm40\zdot\ups09 & $-72\arcd39\arcm35\zdot\arcs3$ & p &   9 &   8 &    9 & H86-140          \\
SMC0095 & 0\uph53\upm42\zdot\ups05 & $-73\arcd21\arcm31\zdot\arcs5$ & p &  30 &  24 &   45 & BS68                      \\
SMC0096$^{(3)}$ & 0\uph53\upm42\zdot\ups31 & $-72\arcd39\arcm14\zdot\arcs6$ & vp &  11 &  10 &   34 & L61-244,B73      \\
SMC0097 & 0\uph54\upm11\zdot\ups00 & $-72\arcd51\arcm54\zdot\arcs1$ & p &  20 &  20 &   22 & BS72                      \\
SMC0098 & 0\uph54\upm46\zdot\ups73 & $-73\arcd13\arcm24\zdot\arcs5$ & c &  36 &  24 &  120 & B80                       \\
SMC0099 & 0\uph54\upm48\zdot\ups24 & $-72\arcd27\arcm57\zdot\arcs8$ & c &  36 &  23 &  137 & B79                       \\
SMC0100 & 0\uph55\upm08\zdot\ups54 & $-72\arcd48\arcm40\zdot\arcs1$ & p &  14 &  14 &   32 & H86-158                   \\
SMC0101 & 0\uph55\upm11\zdot\ups65 & $-73\arcd17\arcm47\zdot\arcs7$ & vp &  24 &  16 &   43 & H86-155                  \\
SMC0102 & 0\uph55\upm12\zdot\ups04 & $-72\arcd41\arcm00\zdot\arcs0$ & p &  33 &  18 &   97 & H86-159                   \\
SMC0103 & 0\uph55\upm29\zdot\ups70 & $-73\arcd04\arcm17\zdot\arcs5$ & c &  31 &  24 &   93 & B83                       \\
SMC0104$^{(4)}$ & 0\uph55\upm32\zdot\ups98 & $-72\arcd49\arcm58\zdot\arcs1$ & c &  37 &  28 &  366 & K34,L53                   \\
SMC0105 & 0\uph55\upm42\zdot\ups99 & $-72\arcd52\arcm48\zdot\arcs4$ & vp &  44 &  29 &   52 & H86-165                  \\
SMC0106 & 0\uph56\upm08\zdot\ups92 & $-73\arcd12\arcm22\zdot\arcs4$ & p &  26 &  18 &   19 & OGLE                      \\
SMC0107$^{(4)}$ & 0\uph56\upm18\zdot\ups68 & $-72\arcd27\arcm50\zdot\arcs4$ & c &  57 &  43 &  874 & NGC330,K35,L54            \\
SMC0108 & 0\uph56\upm34\zdot\ups48 & $-72\arcd30\arcm08\zdot\arcs3$ & p &  15 &  15 &   36 & H86-172                   \\
SMC0109$^{(1)(2)(4)}$ & 0\uph57\upm29\zdot\ups80 & $-72\arcd15\arcm51\zdot\arcs9$ & p &  24 &  18 &  270 & OGLE     \\
SMC0110 & 0\uph57\upm46\zdot\ups14 & $-72\arcd42\arcm20\zdot\arcs9$ & c &  20 &  12 &   36 & H86-178                   \\
SMC0111 & 0\uph57\upm50\zdot\ups23 & $-72\arcd56\arcm36\zdot\arcs9$ & vp &  14 &   9 &   22 & BS88                     \\
SMC0112$^{(3)}$ & 0\uph57\upm57\zdot\ups14 & $-72\arcd26\arcm42\zdot\arcs0$ & vp &  29 &  18 &   43 & H86-179          \\
SMC0113 & 0\uph58\upm16\zdot\ups29 & $-72\arcd38\arcm46\zdot\arcs8$ & p &  24 &  14 &    5 & L61-331     \\
SMC0114 & 0\uph58\upm25\zdot\ups73 & $-72\arcd39\arcm56\zdot\arcs5$ & c &  18 &  12 &   51 & L61-335,      \\
SMC0115 & 0\uph58\upm33\zdot\ups64 & $-72\arcd16\arcm51\zdot\arcs6$ & p &  15 &  15 &   24 & H86-183         \\
SMC0116 & 0\uph59\upm04\zdot\ups54 & $-72\arcd47\arcm11\zdot\arcs7$ & p &  20 &  13 &   21 & OGLE             \\
SMC0117 & 0\uph59\upm13\zdot\ups86 & $-72\arcd36\arcm29\zdot\arcs3$ & c &  45 &  27 &  130 & B96               \\
SMC0118 & 0\uph59\upm48\zdot\ups03 & $-72\arcd20\arcm02\zdot\arcs5$ & c &  41 &  29 &  336 & IC1611,K40,L61      \\
SMC0119$^{(3)}$ & 0\uph59\upm56\zdot\ups87 & $-72\arcd22\arcm24\zdot\arcs4$ & p &   9 &   9 &   22 & H86-186       \\
SMC0120$^{(3)}$ & 1\uph00\upm01\zdot\ups33 & $-72\arcd22\arcm08\zdot\arcs7$ & c &  27 &  21 &  104 & IC1612,K41,L62  \\
SMC0121 & 1\uph00\upm13\zdot\ups03 & $-72\arcd27\arcm43\zdot\arcs8$ & vp &  30 &  24 &   45 & H86-188                  \\
SMC0122$^{(2)}$ & 1\uph00\upm26\zdot\ups77 & $-73\arcd05\arcm11\zdot\arcs6$ & c &  36 &  27 &   44 & B99                \\
SMC0123 & 1\uph00\upm33\zdot\ups09 & $-72\arcd14\arcm23\zdot\arcs0$ & c &  31 &  24 &   78 & H86-189                    \\
SMC0124$^{(4)}$ & 1\uph00\upm34\zdot\ups41 & $-72\arcd21\arcm55\zdot\arcs8$ & c &  34 &  24 &  130 & K42,L63                    \\
SMC0125 & 1\uph00\upm46\zdot\ups75 & $-72\arcd55\arcm41\zdot\arcs4$ & p &  24 &  13 &   20 & OGLE                       \\
SMC0126 & 1\uph01\upm02\zdot\ups01 & $-72\arcd45\arcm05\zdot\arcs2$ & c &  40 &  29 &   95 & L65,H86-192                \\
SMC0127 & 1\uph01\upm17\zdot\ups78 & $-72\arcd13\arcm42\zdot\arcs3$ & c &  36 &  13 &   39 & H86-193                    \\
SMC0128 & 1\uph01\upm37\zdot\ups15 & $-72\arcd24\arcm24\zdot\arcs7$ & c &  36 &  25 &   54 & B105                       \\
SMC0129$^{(1)(4)}$ & 1\uph01\upm45\zdot\ups08 & $-72\arcd33\arcm51\zdot\arcs8$ & c &  29 &  20 &  163 & L66                \\
\hline
}

\setcounter{table}{1}
\MakeTableSep{|l|c|c|c|c|c|c|c|}{12.5cm}{Concluded}{
\hline
\multicolumn{1}{|c|}{Name}&
\multicolumn{1}{|c|}{$\alpha_{2000}$}&
\multicolumn{1}{|c|}{$\delta_{2000}$}&
\multicolumn{1}{|c|}{class}&
\multicolumn{1}{|c|}{$R$}&
\multicolumn{1}{|c|}{$R_{\rm core}$ }&
\multicolumn{1}{|c|}{N} &
\multicolumn{1}{|c|}{Cross-}\\
\multicolumn{1}{|c|} {OGLE-CL-} & & & &
\multicolumn{1}{|c|}{[\arcs]} & {[\arcs]} & &-identifications\\
\hline
SMC0130 & 1\uph01\upm52\zdot\ups23 & $-72\arcd10\arcm57\zdot\arcs6$ & c &  49 &  28 &   79 & B108                      \\
SMC0131 & 1\uph02\upm03\zdot\ups29 & $-72\arcd31\arcm18\zdot\arcs6$ & p &  12 &   8 &   19 & OGLE                      \\
SMC0132 & 1\uph02\upm13\zdot\ups18 & $-72\arcd57\arcm59\zdot\arcs0$ & vp &  20 &  13 &   42 &OGLE                      \\
SMC0133 & 1\uph02\upm30\zdot\ups84 & $-72\arcd19\arcm05\zdot\arcs5$ & c &  24 &  20 &   62 & OGLE                      \\
SMC0134 & 1\uph03\upm11\zdot\ups52 & $-72\arcd16\arcm21\zdot\arcs0$ & c &  28 &  18 &   95 & K47,L70                   \\
SMC0135$^{(1)}$ & 1\uph03\upm16\zdot\ups65 & $-72\arcd44\arcm26\zdot\arcs9$ & p &  43 &  28 &   70 & OGLE              \\
SMC0136$^{(2)}$ & 1\uph03\upm22\zdot\ups05 & $-72\arcd27\arcm57\zdot\arcs0$ & p &  27 &  14 &   12 & OGLE              \\
SMC0137$^{(2)}$ & 1\uph03\upm22\zdot\ups67 & $-72\arcd39\arcm05\zdot\arcs6$ & c &  36 &  18 &   89 & OGLE              \\
SMC0138 & 1\uph03\upm53\zdot\ups02 & $-72\arcd06\arcm10\zdot\arcs5$ & c &  18 &  11 &   35 & L61-420                   \\
SMC0139$^{(1)(4)}$ & 1\uph03\upm53\zdot\ups44 & $-72\arcd49\arcm34\zdot\arcs2$ & c &  20 &  30 &  470 & NGC376,K49,L72    \\
SMC0140 & 1\uph04\upm14\zdot\ups10 & $-72\arcd38\arcm49\zdot\arcs1$ & p &  25 &  18 &   38 & OGLE                      \\
SMC0141 & 1\uph04\upm30\zdot\ups18 & $-72\arcd37\arcm09\zdot\arcs4$ & c &  35 &  24 &   92 & B121                      \\
SMC0142 & 1\uph04\upm36\zdot\ups21 & $-72\arcd09\arcm38\zdot\arcs5$ & c &  41 &  24 &  155 & K50,L74                   \\
SMC0143 & 1\uph04\upm39\zdot\ups61 & $-72\arcd32\arcm59\zdot\arcs7$ & vp &  27 &  23 &   76 & BS125                    \\
SMC0144 & 1\uph04\upm05\zdot\ups23 & $-72\arcd07\arcm14\zdot\arcs6$ & p &  18 &  14 &    6 & OGLE                      \\
SMC0145 & 1\uph05\upm04\zdot\ups30 & $-71\arcd59\arcm24\zdot\arcs8$ & c &  18 &  16 &   86 & L61-439,                  \\
SMC0146 & 1\uph05\upm13\zdot\ups40 & $-71\arcd59\arcm41\zdot\arcs8$ & vp &  14 &  12 &   20 &OGLE                      \\
SMC0147 & 1\uph05\upm07\zdot\ups95 & $-71\arcd59\arcm45\zdot\arcs1$ & c &  22 &  14 &  102 & OGLE                      \\
SMC0148 & 1\uph05\upm09\zdot\ups63 & $-72\arcd36\arcm07\zdot\arcs6$ & p &  28 &  19 &   23 & OGLE                      \\
SMC0149 & 1\uph05\upm21\zdot\ups51 & $-72\arcd02\arcm34\zdot\arcs7$ & c &  36 &  20 &  130 & IC1624,K52,L76            \\
SMC0150 & 1\uph05\upm59\zdot\ups53 & $-72\arcd20\arcm29\zdot\arcs3$ & c &  18 &  12 &   46 & BS131                     \\
SMC0151 & 1\uph06\upm12\zdot\ups62 & $-72\arcd47\arcm38\zdot\arcs7$ & p &  36 &  27 &   20 &OGLE                       \\
SMC0152$^{(2)}$ & 1\uph06\upm21\zdot\ups12 & $-72\arcd49\arcm48\zdot\arcs1$ & p &  20 &   9 &   13 & OGLE              \\
SMC0153$^{(1)}$ & 1\uph06\upm47\zdot\ups74 & $-72\arcd16\arcm24\zdot\arcs5$ & c &  27 &  19 &   91 & K54,L79           \\
SMC0154 & 1\uph07\upm02\zdot\ups27 & $-72\arcd37\arcm18\zdot\arcs2$ & p &  33 &  24 &   70 & B129                      \\
SMC0155 & 1\uph07\upm27\zdot\ups83 & $-72\arcd29\arcm35\zdot\arcs5$ & c &  41 &  28 &   99 & K56                       \\
SMC0156 & 1\uph07\upm28\zdot\ups47 & $-72\arcd46\arcm09\zdot\arcs5$ & c &  41 &  34 &  130 & L80                       \\
SMC0157 & 1\uph07\upm32\zdot\ups36 & $-73\arcd07\arcm11\zdot\arcs3$ & c &  26 &  22 &  100 & K55,L81                   \\
SMC0158$^{(4)}$ & 1\uph07\upm58\zdot\ups97 & $-72\arcd21\arcm19\zdot\arcs5$ & c &  77 &  36 &  985 & NGC416,K59,L83            \\
SMC0159$^{(1)(4)}$ & 1\uph08\upm19\zdot\ups45 & $-72\arcd53\arcm02\zdot\arcs5$ & c & 102 &  73 & 1762 & NGC419,K58,L85    \\
SMC0160 & 1\uph08\upm37\zdot\ups48 & $-72\arcd26\arcm20\zdot\arcs9$ & p &  20 &  15 &   22 & OGLE                      \\
SMC0161 & 1\uph09\upm03\zdot\ups47 & $-73\arcd05\arcm12\zdot\arcs4$ & c &  28 &  15 &   44 & K61                       \\
\hline
}

Table~3 presents the clusters found during visual examination of frames. 
Column 1 shows the proposed name. The equatorial coordinates (J2000) are given 
in columns 2 and 3. Because of small sizes and relatively few members only 
core radii are given in column~4. Crude number of members and
cross-identification are listed in columns 5, and 6,
respectively. Remarks in column 1 are the same as in Table~2.
\setcounter{table}{2}
\MakeTableSep{|l|c|c|c|c|c|}{12cm}{SMC cluster candidates}{
\hline
\multicolumn{1}{|c|}{Name}&
\multicolumn{1}{|c|}{$\alpha_{2000}$}&
\multicolumn{1}{|c|}{$\delta_{2000}$}&
\multicolumn{1}{|c|}{$R$}&
\multicolumn{1}{|c|}{N} &
\multicolumn{1}{|c|}{Cross-}\\
\multicolumn{1}{|c|} {OGLE-CL-} & & 
\multicolumn{1}{|c|} {} & {[\arcs]} & &-identifications \\
\hline
SMC0162 & 0\uph38\upm37\zdot\ups08 & $-73\arcd48\arcm20\zdot\arcs7$ &  13  &   25 & B14            \\                    
SMC0163 & 0\uph38\upm51\zdot\ups28 & $-73\arcd22\arcm27\zdot\arcs1$ &  21  &   21 & HW12           \\                    
SMC0164 & 0\uph38\upm55\zdot\ups55 & $-73\arcd24\arcm31\zdot\arcs7$ &   8  &   15 & OGLE           \\                    
SMC0165 & 0\uph39\upm11\zdot\ups56 & $-73\arcd14\arcm45\zdot\arcs5$ &  12  &   10 & OGLE           \\                    
SMC0166 & 0\uph39\upm11\zdot\ups56 & $-73\arcd14\arcm45\zdot\arcs5$ &   9  &   14 & OGLE           \\                    
SMC0167$^{(2)}$ & 0\uph39\upm25\zdot\ups61 & $-73\arcd06\arcm23\zdot\arcs5$ &   5  &    8 & H86-55 \\                    
SMC0168 & 0\uph39\upm30\zdot\ups62 & $-73\arcd25\arcm25\zdot\arcs8$ &  16  &   11 & OGLE           \\                    
SMC0169 & 0\uph40\upm44\zdot\ups77 & $-73\arcd44\arcm26\zdot\arcs4$ &  16  &   27 & B26            \\                    
SMC0170 & 0\uph40\upm55\zdot\ups45 & $-73\arcd24\arcm06\zdot\arcs6$ &  10  &   21 & OGLE           \\                    
SMC0171$^{(3)}$ & 0\uph41\upm15\zdot\ups30 & $-72\arcd49\arcm54\zdot\arcs9$ &   7  &    6 & B21    \\                    
SMC0172 & 0\uph41\upm48\zdot\ups35 & $-73\arcd23\arcm27\zdot\arcs4$ &   6  &   12 & OGLE           \\                    
SMC0173 & 0\uph42\upm11\zdot\ups62 & $-73\arcd16\arcm02\zdot\arcs1$ &   6  &    6 & OGLE           \\                    
SMC0174 & 0\uph43\upm14\zdot\ups26 & $-73\arcd00\arcm42\zdot\arcs8$ &  21  &   11 & BS17           \\                    
SMC0175 & 0\uph43\upm37\zdot\ups59 & $-72\arcd57\arcm30\zdot\arcs9$ &  12  &   23 & OGLE           \\                    
SMC0176$^{(1)}$ & 0\uph44\upm51\zdot\ups57 & $-73\arcd00\arcm07\zdot\arcs4$ &  14  &   20 & B34    \\                    
SMC0177$^{(3)}$ & 0\uph44\upm55\zdot\ups05 & $-73\arcd10\arcm27\zdot\arcs4$ &   9  &   23 & BS27   \\                    
SMC0178 & 0\uph45\upm10\zdot\ups92 & $-72\arcd52\arcm31\zdot\arcs3$ &   9  &   17 & OGLE           \\                    
SMC0179 & 0\uph45\upm20\zdot\ups88 & $-73\arcd02\arcm07\zdot\arcs7$ &  10  &   15 & OGLE           \\                    
SMC0180 & 0\uph45\upm23\zdot\ups11 & $-72\arcd55\arcm42\zdot\arcs8$ &   7  &   12 & OGLE           \\                    
SMC0181 & 0\uph45\upm27\zdot\ups52 & $-72\arcd53\arcm08\zdot\arcs9$ &  13  &   20 & OGLE           \\                    
SMC0182 & 0\uph46\upm01\zdot\ups63 & $-73\arcd23\arcm44\zdot\arcs4$ &   7  &   10 & H86-76         \\                    
SMC0183 & 0\uph46\upm10\zdot\ups39 & $-73\arcd03\arcm55\zdot\arcs5$ &  10  &   16 & OGLE           \\                    
SMC0184 & 0\uph46\upm11\zdot\ups13 & $-72\arcd49\arcm02\zdot\arcs5$ &   6  &   12 & H86-80         \\                    
SMC0185$^{(3)}$ & 0\uph46\upm34\zdot\ups04 & $-72\arcd45\arcm55\zdot\arcs7$ &   4  &   11 & H86-84 \\                    
SMC0186 & 0\uph46\upm56\zdot\ups26 & $-73\arcd25\arcm24\zdot\arcs7$ &   7  &   15 & H86-85         \\                    
SMC0187 & 0\uph47\upm05\zdot\ups87 & $-73\arcd22\arcm16\zdot\arcs6$ &  14  &   37 & OGLE           \\                    
SMC0188 & 0\uph47\upm24\zdot\ups41 & $-73\arcd12\arcm19\zdot\arcs8$ &  10  &    7 & H86-93         \\                   
SMC0189 & 0\uph48\upm09\zdot\ups12 & $-73\arcd14\arcm19\zdot\arcs1$ &  12  &   14 & OGLE           \\                    
SMC0190$^{(3)}$ & 0\uph48\upm13\zdot\ups20 & $-72\arcd47\arcm34\zdot\arcs7$ &  12  &    7 & H86-99 \\                    
SMC0191$^{(3)}$ & 0\uph48\upm20\zdot\ups19 & $-72\arcd47\arcm42\zdot\arcs1$ &  20  &   27 & H86-100\\                    
SMC0192$^{(1)}$ & 0\uph48\upm26\zdot\ups11 & $-73\arcd00\arcm25\zdot\arcs5$ &  10  &    9 & OGLE   \\                    
SMC0193 & 0\uph48\upm36\zdot\ups55 & $-73\arcd10\arcm45\zdot\arcs2$ &  15  &   26 & OGLE           \\                   
SMC0194$^{(3)}$ & 0\uph49\upm05\zdot\ups58 & $-73\arcd21\arcm09\zdot\arcs8$ &  12  &   13 & BS41   \\                    
SMC0195 & 0\uph49\upm16\zdot\ups45 & $-73\arcd14\arcm56\zdot\arcs8$ &  28  &    4 & BS42           \\                    
SMC0196 & 0\uph49\upm26\zdot\ups81 & $-73\arcd23\arcm55\zdot\arcs2$ &  10  &   11 & OGLE           \\                    
SMC0197 & 0\uph50\upm03\zdot\ups82 & $-73\arcd23\arcm03\zdot\arcs9$ &  24  &   37 & B53            \\                    
SMC0198 & 0\uph50\upm07\zdot\ups51 & $-73\arcd11\arcm25\zdot\arcs9$ &  22  &   60 & H86-112        \\                    
SMC0199 & 0\uph50\upm15\zdot\ups07 & $-73\arcd03\arcm14\zdot\arcs6$ &   7  &    6 & OGLE           \\                    
SMC0200 & 0\uph50\upm38\zdot\ups98 & $-72\arcd58\arcm43\zdot\arcs6$ &  11  &   27 & BS46           \\                    
SMC0201$^{(3)}$ & 0\uph50\upm42\zdot\ups49 & $-73\arcd23\arcm49\zdot\arcs0$ &  19  &   52 & BS48   \\                    
SMC0202 & 0\uph50\upm58\zdot\ups65 & $-73\arcd30\arcm13\zdot\arcs1$ &  10  &   16 & OGLE           \\                    
SMC0203 & 0\uph51\upm10\zdot\ups30 & $-73\arcd35\arcm23\zdot\arcs7$ &   6  &   13 & OGLE           \\                    
SMC0204 & 0\uph51\upm21\zdot\ups33 & $-73\arcd08\arcm18\zdot\arcs7$ &  10  &   31 & H86-121        \\                    
\hline
}

\setcounter{table}{2}
\MakeTableSep{|l|c|c|c|c|c|c|}{12cm}{Concluded}{
\hline
\multicolumn{1}{|c|}{Name}&
\multicolumn{1}{|c|}{$\alpha_{2000}$}&
\multicolumn{1}{|c|}{$\delta_{2000}$}&
\multicolumn{1}{|c|}{$R$}&
\multicolumn{1}{|c|}{N} &
\multicolumn{1}{|c|}{Cross-}\\
\multicolumn{1}{|c|} {OGLE-CL-} & &
\multicolumn{1}{|c|} {} & {[\arcs]} & &-identifications \\
\hline
SMC0205 & 0\uph51\upm31\zdot\ups25 & $-72\arcd58\arcm43\zdot\arcs5$ &  18  &   12 & H86-124       \\                     
SMC0206 & 0\uph51\upm44\zdot\ups34 & $-73\arcd10\arcm01\zdot\arcs3$ &  16  &   23 & H86-123       \\                     
SMC0207 & 0\uph51\upm48\zdot\ups80 & $-72\arcd32\arcm27\zdot\arcs6$ &  12  &   22 & H86-127       \\                     
SMC0208 & 0\uph52\upm02\zdot\ups76 & $-72\arcd49\arcm03\zdot\arcs9$ &   8  &   27 & OGLE          \\                     
SMC0209 & 0\uph52\upm15\zdot\ups25 & $-72\arcd45\arcm58\zdot\arcs1$ &  12  &    8 & OGLE          \\                     
SMC0210$^{(1)}$ & 0\uph52\upm30\zdot\ups30 & $-73\arcd02\arcm59\zdot\arcs0$ &  21  &    5 & B64   \\                     
SMC0211$^{(3)}$ & 0\uph52\upm32\zdot\ups15 & $-73\arcd02\arcm10\zdot\arcs3$ &  14  &    6 & BS57  \\                     
SMC0212$^{(3)}$ & 0\uph52\upm44\zdot\ups52 & $-72\arcd59\arcm24\zdot\arcs2$ &   9  &    8 & H86-134\\                    
SMC0213$^{(3)}$ & 0\uph52\upm48\zdot\ups29 & $-72\arcd59\arcm22\zdot\arcs2$ &  11  &   28 & H86-134\\                    
SMC0214 & 0\uph53\upm08\zdot\ups54 & $-72\arcd49\arcm57\zdot\arcs7$ &   8  &    9 & OGLE           \\                    
SMC0215 & 0\uph53\upm16\zdot\ups84 & $-72\arcd44\arcm02\zdot\arcs8$ &  12  &    7 & OGLE           \\                   
SMC0216 & 0\uph53\upm50\zdot\ups13 & $-72\arcd53\arcm47\zdot\arcs0$ &  40  &   85 & H86-147        \\                    
SMC0217 & 0\uph53\upm56\zdot\ups49 & $-72\arcd51\arcm23\zdot\arcs8$ &   8  &   15 & BS69           \\                    
SMC0218 & 0\uph54\upm22\zdot\ups73 & $-72\arcd41\arcm41\zdot\arcs1$ &  12  &   63 & H86-152        \\                    
SMC0219 & 0\uph54\upm51\zdot\ups52 & $-72\arcd54\arcm59\zdot\arcs0$ &   7  &    7 & B76            \\                    
SMC0220 & 0\uph55\upm14\zdot\ups05 & $-72\arcd36\arcm03\zdot\arcs8$ &  10  &   21 & OGLE           \\                    
SMC0221 & 0\uph55\upm44\zdot\ups75 & $-72\arcd42\arcm18\zdot\arcs2$ &   6  &   23 & OGLE           \\                    
SMC0222 & 0\uph56\upm15\zdot\ups83 & $-72\arcd30\arcm58\zdot\arcs7$ &   8  &   14 & B86            \\                    
SMC0223 & 0\uph56\upm25\zdot\ups59 & $-72\arcd29\arcm45\zdot\arcs1$ &   6  &   12 & BS81           \\                    
SMC0224 & 0\uph56\upm45\zdot\ups88 & $-72\arcd45\arcm06\zdot\arcs0$ &  12  &   17 & OGLE           \\                    
SMC0225 & 0\uph57\upm17\zdot\ups80 & $-72\arcd56\arcm01\zdot\arcs3$ &  14  &   26 & H86-174        \\                    
SMC0226$^{(3)}$ & 0\uph57\upm49\zdot\ups80 & $-72\arcd30\arcm29\zdot\arcs4$ &  14  &   12 & H86-177\\                    
SMC0227$^{(3)}$ & 0\uph57\upm50\zdot\ups23 & $-72\arcd26\arcm23\zdot\arcs6$ &   7  &   15 & H86-175\\                    
SMC0228 & 0\uph58\upm19\zdot\ups42 & $-72\arcd17\arcm56\zdot\arcs6$ &  10  &   21 & H86-181        \\                    
SMC0229 & 0\uph58\upm38\zdot\ups08 & $-72\arcd14\arcm04\zdot\arcs4$ &  20  &    6 & BS272          \\                    
SMC0230 & 1\uph00\upm33\zdot\ups15 & $-72\arcd15\arcm30\zdot\arcs5$ &   9  &   19 & OGLE           \\                    
SMC0231 & 1\uph00\upm58\zdot\ups19 & $-72\arcd32\arcm24\zdot\arcs9$ &  21  &   49 & H86-191        \\                    
SMC0232 & 1\uph01\upm13\zdot\ups58 & $-72\arcd33\arcm03\zdot\arcs5$ &   8  &   13 & OGLE           \\                    
SMC0233 & 1\uph02\upm40\zdot\ups02 & $-72\arcd23\arcm50\zdot\arcs2$ &  12  &   25 & OGLE           \\                    
SMC0234 & 1\uph02\upm53\zdot\ups16 & $-72\arcd24\arcm53\zdot\arcs1$ &  15  &   18 & OGLE           \\                    
SMC0235 & 1\uph03\upm58\zdot\ups96 & $-72\arcd48\arcm18\zdot\arcs3$ &   8  &   20 & OGLE           \\                    
SMC0236 & 1\uph04\upm05\zdot\ups23 & $-72\arcd07\arcm14\zdot\arcs6$ &  12  &   14 & OGLE           \\                    
SMC0237 & 1\uph04\upm22\zdot\ups49 & $-72\arcd50\arcm51\zdot\arcs9$ &  26  &   70 & OGLE           \\                    
SMC0238 & 1\uph07\upm52\zdot\ups23 & $-72\arcd31\arcm54\zdot\arcs9$ &  10  &   20 & OGLE           \\                    
\hline
}

Finding charts were extracted from the "template" {\it I}-band image of each 
field. The charts have size equal to 1.5 of the derived radius of a given 
cluster. To show absolute dimensions of a cluster the bar of 30 arcsec long is 
also given. 

Color-magnitude diagrams of each cluster were constructed based on photometry 
from "Three Color {\it BVI} Maps of the SMC" (Udalski \etal 1998b). We 
included stars located inside the core radii. We did not attempt to perform 
any statistical field subtraction from our CMDs to avoid introducing any 
personal bias in our data. Field subtraction can be performed in many ways by 
potential user of the catalog with the original {\it BVI}-band maps (Udalski 
\etal 1998b) 

The atlas of clusters from the SMC is presented in the Appendix. It contains 
the finding chart with ${V-(B-V)}$ and ${V-(V-I)}$ CMDs of each cluster. 

\Section{Summary}
We present the catalog of clusters found in the central region of the SMC 
(${\approx2.4}$ square degree) observed during the OGLE microlensing survey. 
The catalog contains 238 clusters. 161 of them were found using an automatic 
method. Remaining 77 were detected during visual examination of our CCD 
images. 

We provide the following information for each cluster: equatorial coordinates, 
radii, number of members and cross-identification. Also precise ${V-(B-V)}$ 
and ${V-(V-I)}$ CMDs and finding charts for each cluster are presented. 

The photometric data for clusters from the catalog as well as finding charts 
can be obtain from the OGLE Internet archive: 
{\it http://www.astrouw.edu.pl/\~{}ftp/ogle} and its mirror {\it 
http://www.astro.princeton.edu/\~{}ogle}. 

\Acknow{We would like to thank Dr.\ Eduardo Bica for providing us with the 
computer version of the list of clusters from the SMC. The paper was partly 
supported by the Polish KBN grant 2P03D00814 to A.\ Udalski. Partial support 
for the OGLE project was provided with the NSF grant AST-9530478 to 
B.~Paczy\'nski.}

\newpage
\centerline{\Large\bf Figure Captions}
\vskip1cm
\noindent
Fig.~1. The density images of the scan SMC$\_$SC6 with pixels of 8.2, 12.3 and
20.5 arcsec. North is up and East to the left.

\noindent
Fig.~2. The unsharp masked images of stellar density. The region, scale and 
orientation are the same as in Fig.~1. 
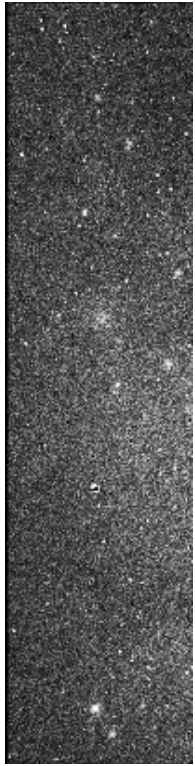
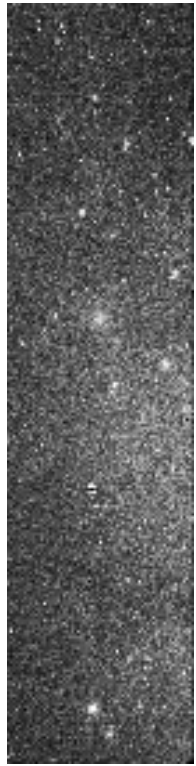
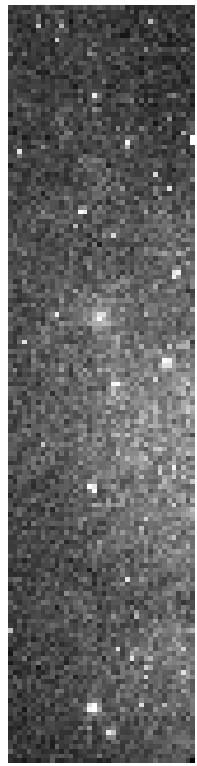
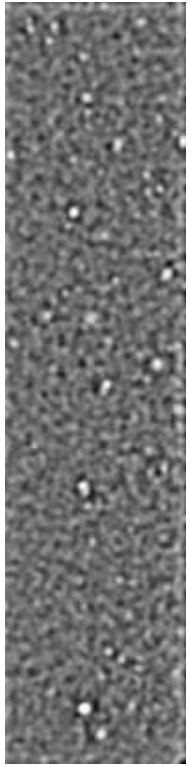
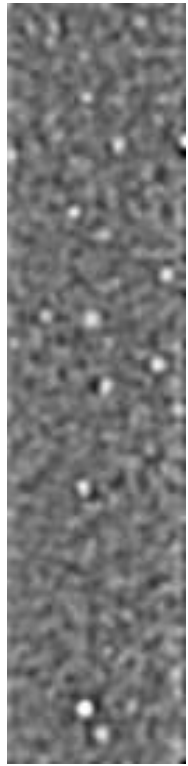
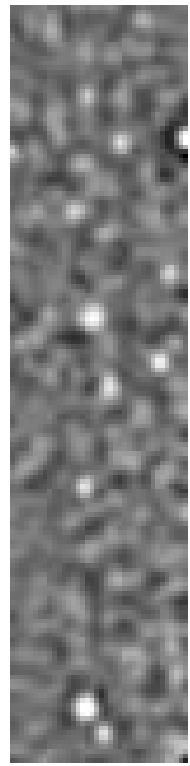
\end{document}